\documentclass{PoS}

\usepackage{fancybox} 
\usepackage{graphicx}
\usepackage{verbatim}
\usepackage{tikz}
\usepackage{amsmath,amsfonts,amssymb}
\def\negcdot{\negmedspace\cdot\negmedspace}

\newcommand*{\chpt}{\raise0.4ex\hbox{$\chi$}PT}
\newcommand*{\schpt}{S\raise0.4ex\hbox{$\chi$}PT}
\newcommand*{\ie}{\textit{i.e.},\ }
\newcommand*{\eg}{\textit{e.g.},\ }
\newcommand*{\vs}{\textit{vs.}\ }

\newcommand*{\et}{\textit{et al.}}

\providecommand*{\prd}[1]{Phys.\ Rev.\ \textbf{D#1}}
\renewcommand*{\prd}[1]{Phys.\ Rev.\ \textbf{D#1}}

\newcommand{\opopo}{\ensuremath{1\!+\!1\!+\!1}}

\newcommand*{\Tr}{\ensuremath{\operatorname{Tr}}}

\newcommand{\trDt}{\ensuremath{\textrm{tr}_{\textrm{\tiny \it D},t}}}

\def\gtwid{{\,\raise.35ex\hbox{$>$\kern-.75em\lower1ex\hbox{$\sim$}}\,}}
\def\ltwid{{\,\raise.35ex\hbox{$<$\kern-.75em\lower1ex\hbox{$\sim$}}\,}}
\def\leftvec{{\raise1.5ex\hbox{$\leftarrow$}\kern-1.00em}}
\def\rightvec{{\raise1.5ex\hbox{$\rightarrow$}\kern-1.00em}}
\def\half{{\scriptstyle \raise.2ex\hbox{${1\over2}$}}}
\def\threehalves{{\scriptstyle \raise.15ex\hbox{${3\over2}$}}}
\def\third{{\scriptstyle \raise.15ex\hbox{${1\over3}$}}}
\def\third{{\scriptstyle \raise.15ex\hbox{${1\over3}$}}}
\def\twothirds{{\scriptstyle \raise.15ex\hbox{${2\over3}$}}}
\def\fourth{{\scriptstyle \raise.15ex\hbox{${1\over4}$}}}

\newcommand{\vslash}{\ensuremath{v\!\!\!\! /}}

\newcommand{\cL}{\ensuremath{\mathcal{L}}}
\newcommand{\cM}{\ensuremath{\mathcal{M}}}

\newcommand{\cO}{\ensuremath{\mathcal{O}}}

\newcommand{\eq}[1]{Eq.~\eqref{eq:#1}}

\def\figref#1{Fig.~\ref{fig:#1}}

\def\eqn#1{\label{eq:#1}}

\title{Staggered Chiral Perturbation Theory for All-Staggered Heavy-Light Mesons}

\ShortTitle{\schpt\ for All-Staggered Heavy-Light Mesons}

\author{\speaker{Javad Komijani} and Claude Bernard\\ 
        Department of Physics, Washington University, St.~Louis, MO 63130, USA\\
        E-mail: \email{jkomijani@wustl.edu}}

\abstract{
In HISQ simulations by the MILC and Fermilab Lattice collaborations, both the light quarks and the charm quark are staggered. 
We extend staggered chiral perturbation theory (\schpt) to include such all-staggered heavy-light mesons.
We assume that the heavy quark action is sufficiently improved that we may take $a m_Q <<1$
(where $m_Q$ is the heavy quark mass),
but also that $m_Q>>\Lambda_{QCD}$ so that a continuum heavy quark expansion is appropriate.
Using this \schpt, the leptonic decay constant of the heavy-light
meson is calculated
 at next-to-leading-order. 
The pattern of taste splittings in the heavy-light meson masses is also investigated.
}

\FullConference{The 30${\rm th}$ International Symposium on Lattice Field Theory -- Lattice 2012\\
                 June 24--29,  2012\\
                 Cairns, Australia}

\begin{document}

\section{Introduction}

\vspace{-2mm}

Heavy-light meson systems provide some of the best ways to test
the Standard Model and look for signs of new physics.  Lattice QCD provides a means of carrying out 
non-perturbative calculations from first principles and with
controlled errors. In setting up a lattice QCD calculation, a key choice is the form of the lattice action
for the quarks. 

Staggered fermions \cite{Kogut:1974ag} are an efficient approach to simulating light quarks.
The ``highly improved staggered quark'' (HISQ) action \cite{HISQ} makes it possible 
to treat charm quarks with the same action as the light quarks.  Thus 
``all-staggered'' simulations of $D$ and $D_s$ mesons are now possible \cite{HISQ-charm}.
There are several advantages to this all-staggered approach. Probably the most
important is that, since heavy and light
quarks have the same action, there are partially conserved heavy-light axial and vector
currents, which therefore need no renormalization.

Lattice computations often involve an extrapolation in light
quark masses to the physical up and down masses, and always require a continuum
extrapolation in lattice spacing.  A version
of chiral perturbation theory (\chpt) that includes the effects of the
discretization errors can help to control these extrapolations. Here, we develop
chiral perturbation theory for all-staggered heavy-light mesons. 

Reference \cite{Aubin:StagHL2006} works out a closely related chiral theory 
for heavy-light mesons with
staggered light quarks but non-staggered heavy quarks (for example,
Fermilab \cite{El-Khadra:1996mp} or NRQCD \cite{NRQCD} quarks).  
In that case, the doubler states of a heavy quark are treated as integrated out, and therefore, heavy-light
mesons have a single taste degree of freedom associated with the light quark.
Here, we need to extend the program developed in Ref.~\cite{Aubin:StagHL2006} 
to include staggered heavy quarks with a taste degree of freedom. 
We assume that the staggered action used (\eg HISQ) 
is improved sufficiently that
we can treat the heavy quark as ``continuum-like,'' with small corrections
from cutoff effects.  We refer to this assumption in short-hand as taking  $am_Q\ll1$, 
where $m_Q$ is mass of the heavy quark. We can then
use the Symanzik effective theory (SET)  \cite{SET} to describe
the discretization effects on the heavy quarks, as well as on the light quarks.

In the continuum limit, there is an exact $SU(4)$ symmetry acting
on tastes; this symmetry is broken at $\cO(a^2)$ in the lattice spacing $a$. 
The corresponding discretization errors in the light-light sector split the masses of
mesons with different tastes, which may be 
understood using staggered chiral perturbation theory (\schpt) \cite{LEE_SHARPE,SCHPT}.
For typical values of $a^2$, 
the taste splittings of the light pseudoscalar
mesons can be comparable to the masses themselves. Schematically, we say $a^2\sim m^2_\pi$,
where appropriate powers of $\Lambda_{QCD}$ or $\Lambda_{\chi}$ (the chiral scale) are 
implicitly inserted to match the dimensions in such comparisons.  
The taste splittings are therefore
included in the leading order (LO) light-light Lagrangian, which is of ${\cO}(m^2_\pi)$.

For heavy-light mesons composed of staggered quarks,
the situation is different. The LO Lagrangian in the continuum is of ${\cO}(k)$,
where $k$ is the residual momentum of the heavy-light meson, and we 
assume that $k\sim m_\pi$. Thus, it is reasonable
to treat taste violations, which are of ${\cO}(a^2)$,
as next-to-leading order (NLO) corrections, and that is what we do here. 
The LO heavy-light Lagrangian is then taste invariant. 
This power counting can be checked with HISQ simulations, where the splittings in {\em squared}\/ meson 
masses remain roughly constant as one increases a valence quark mass from the light
quark regime to the charm regime \cite{MILC-HISQ-CONFIGS}. (See also
\figref{MediumCoarseMILCensemble} below.)  Therefore the splittings for the masses themselves
are much smaller for heavy-light mesons than for light mesons.
For example, at $a\approx0.12\;$fm
the measured \cite{MILC-HISQ-CONFIGS} 
taste splitting between the root-mean-squared (RMS) $D_s$ meson and the lightest 
$D_s$ meson is only about 11 MeV, while it is 
about 110 MeV for the pion (80\% of $m_\pi$). 

\vspace{-1mm}

\section{The \schpt\ Lagrangian for Heavy-Light Mesons}
\label{sec:ls-lag}

\vspace{-2mm}

We first discuss a heavy-light meson where the light quark has a taste degree of freedom,
but the heavy quark does not \cite{Aubin:StagHL2006}.
Due to the heavy quark spin symmetry in the static limit, the heavy vector
and pseudoscalar mesons are incorporated into the following field,
which destroys a heavy-light meson,
\begin{equation}
  H_a = \frac{1 + \vslash}{2}\left[ \gamma^\mu B^{*}_{\mu a}
    + i \gamma_5 B_{a}\right]\ ,
\end{equation}
where $v$ is the meson's velocity, and $a$ is the combined light quark flavor-taste index. 
Note that we use $B$ to
denote a generic pseudoscalar heavy meson and $B^*$ to denote the corresponding
vector meson, but the practical application of this calculation will, at least in the
first instance, be to $D$ mesons.

Now, we assume the heavy quark is also implemented by a staggered fermion.  
Then we generalize the definition of the annihilation operator of a heavy-light meson as
\begin{equation} \label{eq:H_definition}
  H_{\alpha a} = \frac{1 + \vslash}{2}\left[ \gamma^\mu B^{*}_{\mu \alpha a}
    + i \gamma_5 B_{\alpha a}\right]\ ,
\end{equation}
where $v$ is the meson's velocity, $\alpha$ is the heavy-quark taste index,
and $a$ is the combined flavor-taste index of the light quark. 
The conjugate field creates a heavy-light meson
\begin{equation} \label{eq:Hbar_definition}
  \overline{H}_{a\alpha} \equiv \gamma_0 H^{\dagger}_{a\alpha}\gamma_0 =
  \left[ \gamma^\mu B^{\dagger *}_{\mu a\alpha}
    + i \gamma_5 B^{\dagger}_{a\alpha}\right]\frac{1 + \vslash}{2}\ .
\end{equation}

So far $H$ is treated as a $4 \times 4n$ matrix in the taste and the flavor space of quarks. 
Instead of attaching separate indices to the tastes of the light and the heavy quarks, 
one might use just one index as the taste of the meson. Then, one treats $H$ as a $n$-component
vector in the flavor
space of the light quark, while each 
element ($H_i$) is a $4\times 4$ matrix in the taste space of the meson. 
To implement this approach, the light flavor-taste index is first traded for 
a pair of indices representing flavor and taste separately.
We use Latin indices in the middle of the alphabet $(i, j, ...)$ as pure flavor indices. 
We can then combine the tastes of the heavy and the light quarks and use an index such 
as $\Xi$ (where
$\Xi=1,\dots16$) as the taste of the meson.
Therefore, the $i$th element of the field destroying a heavy-light meson in the light flavor space can be represented by 
$H_i = \sum_{\Xi=1}^{16} H_{i \Xi } T_\Xi$ and
its conjugate by $\overline{H}_i = \sum_{\Xi=1}^{16} \overline{H}_{i\Xi} T_\Xi$,
with the Hermitian taste generators 
$T_\Xi \in \{ \xi_5, i\xi_{\mu 5}, i\xi_{\mu\nu} , \xi_{\mu}, \xi_I\}$.

The leading-order staggered version of the heavy-light meson Lagrangian then 
looks the same as in the case with only light staggered fields, or indeed in standard
continuum \chpt, with the only change being the extra taste degrees of freedom in the
fields.
The LO chiral Lagrangian involving the heavy-light meson fields is:
\begin{equation}
  \cL_1  =  -i \Tr(\overline{H} H v\negcdot \leftvec D )
  + g_\pi \Tr(\overline{H}H\gamma^{\mu}\gamma_5 
  \mathbb{A}_{\mu})\ ,
\end{equation}
where $ (\overline{H}H)_{ab} \equiv \overline{H}_{a \alpha} H_{\alpha b},$ and
$\Tr$, here and below, 
means the complete trace over flavor and taste indices and, where relevant,
Dirac indices.  
This Lagrangian has separate $SU(4)$ taste symmetries on the heavy and light quarks, as well as spin symmetry of the heavy quark.

To derive the heavy-light decay constants, we also will need 
the chiral representative of the axial heavy-light current.
Alternatively, one can work with the left-handed current.
The left-handed current that destroys a heavy-light meson
of taste $\Xi$ and light flavor $i$ is $j^{\mu,i \Xi}$, which at LO takes the form  
\begin{equation}\label{eq:LOcurrent}
  j^{\mu,i \Xi}_{\rm LO} = \frac{\kappa}{2}\; 
  \trDt\bigl(T_\Xi\gamma^\mu\left(1-\gamma_5\right)H\sigma^\dagger\lambda^{(i)} \bigr) \ ,
\end{equation}
where $\trDt$ is a trace over Dirac and taste indices, and  $\lambda^{(i)}$ is a constant row vector that fixes the flavor of the light quark:
$(\lambda^{(i)})_j = \delta_{i j}$.   
The decay constant $f_{B_{i \Xi}}$ is defined by the matrix
element
\begin{equation}\label{eq:matrix_element}
  \left\langle 0 \left| j^{\mu,i' \Xi'}
  \right| B_{i \Xi}(v) \right\rangle =i
  f_{B_{i \Xi}}\sqrt{m_{B_{i \Xi}}} v^{\mu} \delta_{\Xi \Xi'}\delta_{ii'}\ ,
\end{equation}
where the state $|B_{i \Xi}(v) \rangle $ is normalized non-relativistically,
corresponding to our non-relativistically normalized heavy meson field $B$.
At LO in the heavy-light chiral theory,
$j^{\mu,i' \Xi'}_{\rm LO} = i\kappa v^\mu {B_{i' \Xi'}}$, which gives 
$f_{B_{i \Xi}}^{\rm LO} = \kappa/\sqrt{m_{B_{i \Xi}}}$.

To go beyond leading order, we must encode the discretization errors in the
chiral theory.  To do this we consider a series of effective field theories.
Assuming $am_Q << 1$, we can derive 
a Symanzik effective theory (SET) from the staggered lattice Lagrangian.
One can then use the fact that 
$m_Q$ is large compared to $\Lambda_{QCD}$ to organize heavy quark effects
with  heavy quark effective theory (HQET). 
Finally, when residual momenta and light quark masses are small compared to the chiral
scale $\Lambda_\chi\sim 1\;$GeV, the physics of light-light and heavy-light mesons
may be described by a chiral effective theory. Here, we assume the power counting $p_\pi^2\sim m^2_\pi\sim m_q\sim a^2$ for the light mesons
as in Ref.~\cite{Aubin:StagHL2006}. 

There are three different types of 4-quark operators in the SET at order $a^2$:
\begin{eqnarray}
  a^2\cO_{ss'tt'} = & c_1 & a^2\overline{q}_l(\gamma_s\otimes\xi_t)q_l 
			      \overline{q}_{l'}(\gamma_{s'}\otimes\xi_{t'})q_{l'} \nonumber \\ 
		  + & c_2 & a^2\overline{q}_l(\gamma_s\otimes\xi_t)q_l 
				  \overline{q}_h(\gamma_{s'}\otimes\xi_{t'})q_h \nonumber \\ 
		  + & c_3 & a^2\overline{q}_h(\gamma_s\otimes\xi_t)q_h 
				  \overline{q}_{h'}(\gamma_{s'}\otimes\xi_{t'})q_{h'}  \ ,
\eqn{4qops}
\end{eqnarray}
where $q_l$ and $q_h$ are the light and heavy quark fields,
and $\gamma_s$ and $\xi_t$ are any of the 16 spin or taste matrices, respectively.
The operators listed in \eq{4qops} are generic, and stand for a linear combination
of all operators with the same heavy and light quark structure.
The first type of operator has only light quarks and hence
respects the heavy taste symmetry.  Its contributions to the NLO chiral
Lagrangian are thus essentially
the same as those that appear when the heavy quark has no taste, \ie those of 
Ref.~\cite{Aubin:StagHL2006}.
Although the third type breaks the heavy taste symmetry, 
it contributes only in trivial ways to the heavy-light Lagrangian, which by definition 
has only two meson fields  and does not  describe heavy-light meson scattering.
New chiral terms arise only from
from the second type of operator, which has both heavy and light fields.

Using  a spurion analysis, we find that the operators of the second type lead to the following two 
contributions to the mass term at NLO Lagrangian:
\begin{eqnarray}
  \cL_{2,a^2}^{A2} & = & K_{A1}a^2 \Tr\left(\overline{H} \xi_{5   } H \xi_{    5} \right) +
			K_{A2}a^2 \Tr\left(\overline{H} \xi_{ \mu} H \xi_{\mu  } \right) +
			K_{A3}a^2 \Tr\left(\overline{H} \xi_{5\mu} H \xi_{\mu 5} \right) \nonumber \\*&&{}+
			K_{A4}a^2 \Tr\left(\overline{H} \xi_{\mu\nu} H \xi_{\nu\mu} \right) +
			K_{A5}a^2 \Tr\left(\overline{H} \gamma_{5\mu}          H  \gamma^{\mu 5}          \right) \nonumber \\*&&{}+
			K_{A6}a^2 \Tr\left(\overline{H} \gamma_{5\mu} \xi_{5 } H  \gamma^{\mu 5} \xi_{ 5} \right)  +
			K_{A7}a^2 \Tr\left(\overline{H} \gamma_{\mu\nu} \xi_{\lambda } H  \gamma^{\nu\mu} \xi_{\lambda} \right) \nonumber \\*&&{}+
			K_{A8}a^2 \Tr\left(\overline{H} \gamma_{\mu\nu} \xi_{5\lambda } H  \gamma^{\nu\mu} \xi_{\lambda 5} \right) +
			K_{A9}a^2 \Tr\left(\overline{H} \gamma_{5\mu} \xi_{\nu\lambda} H  \gamma^{\mu 5} \xi_{\lambda\nu} \right) \ ,
\eqn{typeA}
\end{eqnarray}
and
\begin{eqnarray}
  \cL_{2,a^2}^{B2} & = & \sum_\mu \Bigl[K_{B1}a^2 \Tr\left(\overline{H} \gamma_{\nu\mu} \xi_{\mu }   H  \gamma^{\mu\nu} \xi_{\mu }  \right) +
			K_{B2}a^2 \Tr\left(\overline{H} \gamma_{\nu\mu} \xi_{5\mu } H  \gamma^{\mu\nu} \xi_{\mu 5} \right)  \nonumber \\*&&{}+
			K_{B3}a^2 v^\mu v_\mu \Tr\left(\overline{H} \xi_{\nu\mu} H \xi_{\mu\nu} \right) +
			K_{B4}a^2 \Tr\left(\overline{H} \gamma_{5\mu} \xi_{\nu\mu } H  \gamma^{\mu 5} \xi_{\mu\nu} \right) \Bigr] \ ,
\eqn{typeB}
\end{eqnarray}
where $\cL_{2,a^2}^{A2}$ (the ``type-A'' contribution) results from operators that are 
invariant over the full Euclidean space-time rotation group, $SO(4)$,
as well as a corresponding $SO(4)$ of taste, and $\cL_{2,a^2}^{B2}$ 
(the ``type-B'' contribution) results from operators 
that couple spin and taste and break these $SO(4)$ symmetries.

\vspace{-1mm}

\section{Results and Conclusions}
\label{sec:Mass}

\vspace{-2mm}

The $\cO(a^2)$  contributions to the Lagrangian, $\cL_{2,a^2}^{A2}$ and $\cL_{2,a^2}^{B2}$,
give different NLO mass corrections to different tastes of the heavy-light mesons.
The type-A terms split the masses into the five $SO(4)$ taste multiplets, 
labeled by I, V, T, A, and P.
The type-B terms  split these representations and give different masses 
to the time and spatial components, such as $\xi_0$ and $\xi_i$ for the vector
taste (V) representation. Note that no splitting, either between $SO(4)$ multiplets,
or within the multiplets, comes from the one-loop diagrams.  This
can be proved using 
the exact $SU(4)$ heavy-quark taste symmetry of the LO Lagrangian and
the discrete taste symmetry, which corresponds
to the shift symmetry in the staggered action \cite{Aubin:StagHL2006}. 

Based on the pattern of splittings seen at LO  in light mesons, 
the 4-quark operators with the
taste structure $\xi_{\mu5}$ and spin $I$, $\gamma_5$, or $\gamma_{\mu\nu}$ 
appear to be dominant. These operators
contribute to the $C_4$ chiral operator \cite{SCHPT}, which gives the characteristic 
P, A, T, V, I (lowest-to-highest) ordering of squared masses, with roughly equal spacing.
Note that, for light mesons, only type-A 4-quark operators are relevant to LO, because type-B 
operators have no chiral representatives until NLO \cite{LEE_SHARPE}.
Here the corresponding type-A 4-quark operators give rise to the chiral operators 
  $K_{A3}a^2 \Tr(\overline{H} \xi_{5\mu} H \xi_{\mu 5})$ and
  $K_{A8}a^2 \Tr(\overline{H} \gamma_{\mu\nu} 
\xi_{5\lambda } H  \gamma^{\nu\mu} \xi_{\lambda 5} )$. These produce the same equal-spacing 
pattern
for the $SO(4)$ taste representations of heavy-light mesons:  
See Table~\ref{tab:TasteSplittingA}.  For type-B operators, one might guess that the 
taste $\xi_{\mu5}$ and spin  $\gamma_{\mu\nu}$ 4-quark operator would be dominant, since
it is the only type-B operator that has the same spin and taste as one of the dominant
type-A operators. This 4-quark operator gives rise to the chiral operator
  $\sum_\mu  K_{B2}a^2 \Tr(\overline{H} 
\gamma_{\nu\mu} \xi_{5\mu } H  \gamma^{\mu\nu} \xi_{\mu 5})$.
Table~\ref{tab:TasteSplittingB} shows the pattern of mass splitting that stems from this 
type-B operator. 

\begin{figure}[thbp]
\begin{center}
\includegraphics[width=4in]{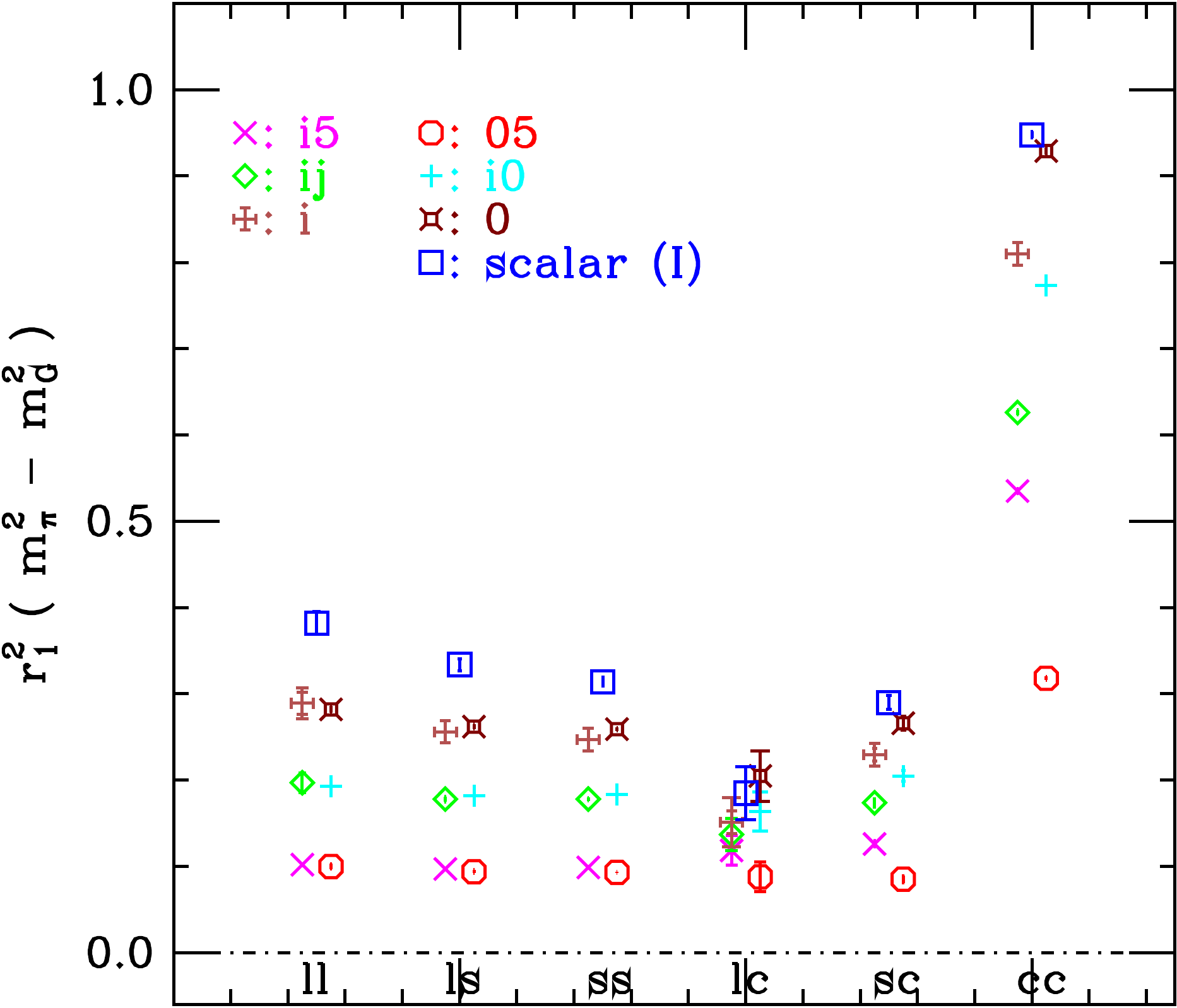}
\end{center}
\caption{MILC HISQ ensemble at $a\approx0.15\;$fm and $m_l=0.2m_s$ \cite{MILC-HISQ-CONFIGS}.
Squared mass splitting between pions of different tastes and the Goldstone pion in units of
$r_1$.  The types of quarks in the mesons are shown on the abscissa: l, s, and c stand for
light (u,d), strange, and charm quarks, respectively.}
\label{fig:MediumCoarseMILCensemble}
\end{figure}

The patterns given in Tables~\ref{tab:TasteSplittingA} and \ref{tab:TasteSplittingB} 
are qualitatively present in the MILC data, shown
in \figref{MediumCoarseMILCensemble}.  Note in particular the ``sc'' case, for which the
errors are small enough that the pattern of $SO(4)$ breaking for heavy-light mesons
is clear.  It is non-trivial that the time component of taste is higher than the space
component in two cases ($\xi_0$ \vs $\xi_i$ and $\xi_{i0}$  \vs $\xi_{ij}$) but not
in the third case ($\xi_{05}$  \vs $\xi_{i5}$), just as in Table~\ref{tab:TasteSplittingB}.  
Although the chiral theory is
not applicable to the ``cc'' case, it is interesting to see that the structure
that would correspond to the dominant type-B operator 
gets particularly strong there, with near degeneracies of between members of different
$SO(4)$ multiplets, in particular $\xi_0$ and $I$, or $\xi_{i0}$ and $\xi_i$.
\begin{table}
\caption{Taste splittings due to the apparently dominant type-A operators.}
\label{tab:TasteSplittingA}
\begin{center}
\begin{tabular}{|c||c|c|c|c|c|} \hline
{} & $\triangle_{m_Q}(\xi_5)$ & $\triangle_{m_Q}(\xi_{5\mu})$  & $\triangle_{m_Q}(\xi_{\mu\nu})$ 
& $\triangle_{m_Q}(\xi_\mu)$ &  $\triangle_{m_Q}(I)$ \\ \hline \hline
  $8 a^2 (K_{A3}+4K_{A8})$  &  -4  &  -2 &  0 &  +2 &  +4 \\ \hline 
\end{tabular}
\end{center}
\end{table}
\begin{table}
\caption{Taste splittings due to the apparently dominant type-B operator.}
\label{tab:TasteSplittingB}
\begin{center}
 \begin{tabular}{|c||c|c c|c c|c c|c|} \hline
{} & $\triangle_{m_Q}(\xi_5)$ & $\triangle_{m_Q}(\xi_{05})$ & $\triangle_{m_Q}(\xi_{i5})$ & $\triangle_{m_Q}(\xi_{ij})$ 
& $\triangle_{m_Q}(\xi_{i0})$ & $\triangle_{m_Q}(\xi_i)$ & $\triangle_{m_Q}(\xi_0)$ & $\triangle_{m_Q}(I)$ \\ \hline \hline
  $8 a^2 K_{B2}$  &-6 &-6 &-2 &-2 &+2 &+2 &+6 & +6\\ \hline 
\end{tabular}
\end{center}
\end{table}

We can also calculate the decay constant of the heavy-light meson to NLO. 
At one loop, we express the decay constant as
\begin{equation}\label{eq:fB}
        f_{B_{x\Xi}} = f_{B_{x}}^{\rm LO} \left( 1 + \frac{1}{16\pi^2 f^2}\;
        \delta\! f_{B_{x}} + {\rm analytic\ terms} \right) \ .
\end{equation}
The lowest order term $f_{B_{x}}^{\rm LO}$ depends only on the light valence
flavor as $f_{B_{x}}^{\rm LO} = \kappa/\sqrt{m_{B_{x}}}$ because the 
taste splittings of heavy-light masses appear at NLO. 
We can divide the terms contributing to the decay constant at NLO into two parts. 
The first part comes from the operators which obey the $SU(4)$ taste symmetry of 
the heavy quarks, while
the second stems from those breaking the heavy taste symmetry. 
The first part contributes both to the analytic terms and 
to the chiral logarithms, $\delta\! f_{B_{x}}$ in \eq{fB}.  It 
is independent of the taste of the meson, and indeed is the same as
the result in Ref.~\cite{Aubin:StagHL2006}, because it obeys 
both the residual discrete taste symmetry
and the $SU(4)$ taste symmetry of the heavy quarks at LO.
The second part contributes only to the analytic terms in \eq{fB},
which depend on the taste of the meson.   

We find the decay constant of the heavy-light meson in the partially quenched case is:
\begin{eqnarray}\label{eq:1p1p1_pq_fB}
  \left(\frac{f_{B_{x\Xi}}}{f_{B_{x}}^{\rm LO}}\right)_{\opopo} 
   &= & 1 + \frac{1}{16\pi^2f^2}
      \frac{1+3g_\pi^2}{2}
      \Biggl\{-\frac{1}{16}\sum_{f,\Xi'} \ell(m_{xf,\Xi'}^2)  \nonumber \\*&&{}-
	\frac{1}{3}\sum_{j\in \cM_I^{(3,x)}} 
	\frac{\partial}{\partial m^2_{X,I}}\left[ R^{[3,3]}_{j}(
	  \cM_I^{(3,x)};  \mu^{(3)}_I)\ell(m_{j}^2) \right] \nonumber \\*&&{}- 
	\biggl( a^2\delta'_V \sum_{j\in \cM_V^{(4,x)}}
	\frac{\partial}{\partial m^2_{X,V}}
	\left[ R^{[4,3]}_{j}( \cM_V^{(4,x)}; \mu^{(3)}_V )
	\ell(m_{j}^2) \right]
	    + [V\to A] \biggr)
      \Biggr\}  \nonumber \\*&&{}+
       c_s (m_u + m_d + m_s) + c_v m_x + c_{a,\Xi} a^2 \ ,
\end{eqnarray}
where $x$ is the valence flavor, $\Xi$ is the valence taste,
$f$ runs over the three sea quarks $u$, $d$, and $s$,  
$\Xi'$ runs over the 16 meson tastes, and other notation 
is explained in Ref.~\cite{Aubin:StagHL2006}.
In \eq{1p1p1_pq_fB}, $c_{a,\Xi}$ is the only coefficient which depends on the taste of the heavy meson.

In summary, we have generalized the chiral Lagrangian of a heavy-light meson to the case where
both heavy and light quarks have taste degrees of freedom.
We have obtained the NLO chiral Lagrangian, which breaks the $SU(4)$ heavy taste symmetry.
Moreover we have derived the NLO decay constants and the taste splittings 
of the heavy-light masses,
and have used them to understand, qualitatively, the pattern of splittings seen in the
heavy-light HISQ data.

We thank our colleagues in MILC for Figure 1, and Andreas Kronfeld 
for helpful discussions on the use of effective theories for this
problem.
Our work was supported in part by the U.S. Department of Energy under Grant DE-FG02-91ER40628.

\end{document}